\documentclass[letterpaper, 10 pt, conference]{ieeeconf}  
\IEEEoverridecommandlockouts                              
\usepackage[absolute]{textpos} 
\usepackage{threeparttable}  
\usepackage{siunitx}
\usepackage{booktabs}
\usepackage{graphicx} 
\graphicspath{ {./images/} }
\usepackage{array}
\usepackage[table]{xcolor}
\definecolor{ao(english)}{rgb}{0.0, 0.5, 0.0}
\definecolor{orange(webcolor)}{rgb}{1.0, 0.65, 0.0}
\definecolor{ao}{rgb}{0.0, 0.0, 1.0}
\definecolor{aqua}{rgb}{0.0, 1.0, 1.0}
\definecolor{patriarch}{rgb}{0.5, 0.0, 0.5}
\definecolor{brown(web)}{rgb}{0.65, 0.16, 0.16}
\newcolumntype{L}{>{\centering\arraybackslash}m{3cm}}

\title{\LARGE \bf
Energy-efficient Blood Pressure Monitoring based on Single-site Photoplethysmogram on Wearable Devices
}
\author{Wenrui Lin$^{1}$$^{*}$, Berken Utku Demirel$^{1}$, Mohammad Abdullah Al Faruque$^{1}$, and G.P. Li$^{1}$
\thanks{$^{1}$All the authors are with the Department of Electrical Engineering and Computer Science, University of California, Irvine, CA 92617, USA.
$^{*}$Email: 
        {\tt\small wenruil3@uci.edu}}%
}

\begin{document}

\setlength{\TPHorizModule}{\paperwidth}
\setlength{\TPVertModule}{\paperheight}
\begin{textblock}{0.8}[0.5,0.5](0.5,0.96)
        \noindent\small{{\copyright 2021 IEEE. Personal use of this material is permitted. However, permission from IEEE must be obtained for all other uses, in any current or future media, including reprinting/republishing this material for advertising or promotional purposes, creating new collective works, for resale or redistribution to servers or lists, or reuse of any copyrighted component of this work in other works. Published in the 2021 43nd Annual International Conference of the IEEE Engineering in Medicine and Biology Society (EMBC).}}
\end{textblock}

\maketitle
\thispagestyle{empty}
\pagestyle{empty}

\begin{abstract}
The paper proposes accurate Blood Pressure Monitoring (BPM) based on a single-site Photoplethysmographic (PPG) sensor and provides an energy-efficient solution on edge cuffless wearable devices. Continuous PPG signal preprocessed and used as input of the Artificial Neural Network (ANN), and outputs systolic BP (SBP), diastolic BP (DBP), and mean arterial BP (MAP) values for each heartbeat. The improvement of the BPM accuracy is obtained by removing outliers in the preprocessing step and the whole-based inputs compared to parameter-based inputs extracted from the PPG signal. Performance obtained is 3.42 ± 5.42 mmHg (MAE ± RMSD) for SBP, 1.92 ± 3.29 mmHg for DBP, and 2.21 ± 3.50 mmHg for MAP which is competitive compared to other studies. This is the first BPM solution with edge computing artificial intelligence as we have learned so far. Evaluation experiments on real hardware show that the solution takes 42.2 ms, 18.2 KB RAM, and 2.1 mJ average energy per reading.
\newline

\end{abstract}

\section{INTRODUCTION}

Continuous blood pressure monitoring (BPM) is important for prevention and early diagnosis of cardiovascular diseases [1]. Current standard ambulatory BPM devices use an oscillometric cuff-based method which can cause physical discomfort with repeated inflations and deflations [2]. Extensive studies are seeking cuff-less, unobtrusive monitoring solutions. Based on photoplethsymogram (PPG) and/or electrocardiogram (ECG), there are two main cuffless BP estimation methods: pulse transit time (PTT) based; and feature extraction of the PPG waveform [3]. PPG is an optical signal related to peripheral blood volume pulsations and its waveform has been proven to have a good correlation with BP [4]. It has the huge advantage of low cost and small form. However, obtaining an accurate estimation of BP using single PPG sensors is challenging, especially for reflective type [5]. Parameter-based BPM technique typically extracts features from the PPG in the time and frequency domain. However, the whole-based technique, which uses raw signal for BPM, usually has advantages in terms of accuracy [6]. The daily use of wearable devices generates an enormous amount of raw data. Sending them over Bluetooth requires a huge amount of energy and introduces latency. To overcome these limitations, edge computing where all the processing is done on wearable devices is introduced [7].

To achieve all those goals, we propose an energy-efficient BPM solution to enable long-term BPM on wearable devices, based on Artificial Neural Network (ANN) with single-site PPG waveform input. Evaluation on real hardware shows that our solution is fast, low-memory consumed, and energy-efficient while providing accurate BPM on an edge device.

\begin{figure*}[!htbp]
    \centering
    \includegraphics[width=0.88\textwidth]{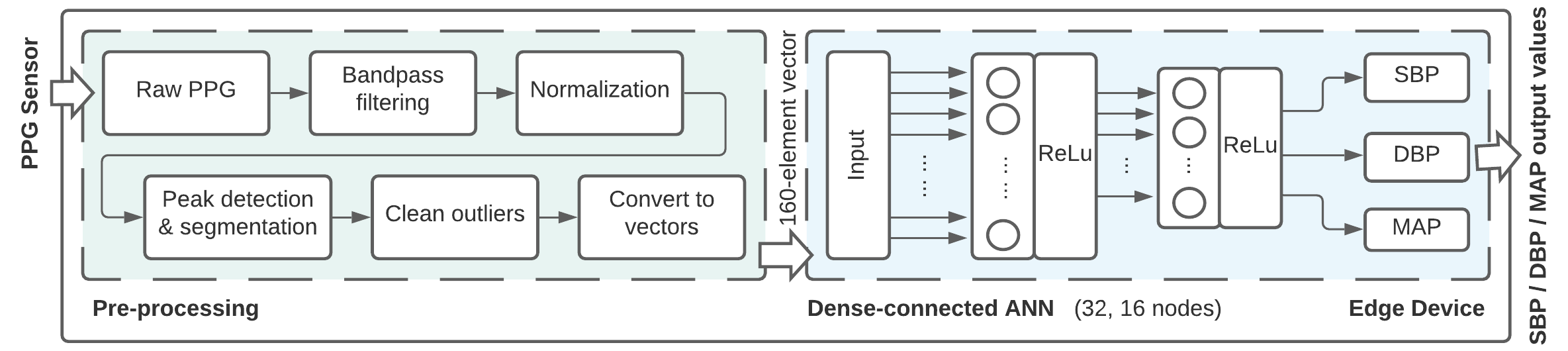}
    \caption{Energy-efficient BPM solution on edge device}
\end{figure*}

\section{RELATED WORK}

Complete BP waveform can be predicted from an ECG and/or PPG time series with a nonlinear autoregressive model with exogenous inputs implemented by ANN [3] and was validated on the 15 subjects from the MIMIC-II database. Heartbeats were removed if the amplitude was outside ± 5 standard deviations to clean outliers. Another BP waveform reconstruction study is validated in the MIMIC database from 90 subjects and extracts SBP and DBP from the reconstructed BP waveform [4]. However, we believe BP waveform prediction is redundant for a wearable device aimed at BP values reading.

Whole-based BP estimation is proposed in [6] with machine learning algorithms to predict the BP values. This study used 1323 records (each 15s) belonging to 441 individuals extracted from the MIMIC-II database and the performance was close to the standard AAMI limits. More pervasively, parameter-based BPM is studied. A study that extracts 21 parameters from each heartbeat (total 15k pulsations) of PPG and feeds forward ANN with two hidden layers shows better accuracy than the linear regression method [8].

Authors in [9] introduce a novel convolutional neural network (CNN) for BPM with whole-based ECG and PPG signals. Records of 604 subjects are cleaned and selected from the MIMIC-II database, and around 200 segments per subject for 10s were used to evaluate their model. However, these deep learning methods are expensive in terms of time, memory, and energy, thus, are unsuitable for wearable devices with edge computing.

\section{PROPOSED METHODOLOGY}
In this section, we detailly introduce our BPM solution from a continuous PPG signal. Preprocessing steps and a nonlinear regression with 2-layer ANN which could converge to embedded electronics is depicted in Fig.1. 

\subsection{Database}
 
We use MIMIC-III Waveform Database from PhysioNet [10], which contains ICU records including PPG, arterial blood pressure (ABP) waveforms (sampling at 125Hz). 11 subjects are randomly chosen from recordings that consist of 10 hours of the PPG and ABP signals.

We extracted more than 40k heartbeat PPG samples with corresponding BP values per subject. BP values were extracted from the ABP signal for each heartbeat. Therefore, there are extra steps to process ABP signals which are not implemented in the embedded system. The SBP and DBP values are assigned respectively as the maximum and minimum ABP amplitude in each heartbeat. The MAP value is calculated by expression: $MAP = (DBP * 2 + SBP)/3$. To suppress the influence of outliers in BP labels, we discard heartbeats with BP values outside the range of ± 5 standard deviations (STDs) from mean. No manual discard of any heartbeat sample in this study.
The distribution of BP values extracted for all selected subjects are shown in Table I:

\begin{table}[!htbp]
\caption{Description of BP values extracted}
\centering
\begin{tabular}{|c||c|c|c|c|c|c|}
\hline
&  \multicolumn{2}{c|}{SBP (mmHg)} & \multicolumn{2}{c|}{DBP (mmHg)} & \multicolumn{2}{c|}{MAP (mmHg)} \\
\hline
Subject  & Mean &  STD  & Mean & STD & Mean & STD \\
\hline\hline
3000397 & 103.7 & 9.7 & 57.6 & 3.2 & 73.0 & 5.1 \\
\hline
3001689 & 136.4 & 11.5 & 64.4 & 4.5 & 88.4 & 5.9\\
\hline
3002090 & 93.2 & 6.3 & 42.4 & 3.8 & 59.3 & 3.9\\
\hline
3006202 & 115.4 & 15.1 & 54.0 & 4.0 & 74.5 & 7.6\\
\hline
3011085 & 118.4 & 6.3 & 78.5 & 3.7 & 91.8 & 4.2\\
\hline
3011536 & 96.4 & 8.9 & 54.7 & 4.3 & 68.6 & 5.7\\
\hline
3505162 & 150.9 & 26.8 & 76.1 & 9.4 & 101.0 & 14.7\\
\hline
3505469 & 91.3 & 5.5 & 54.8 & 4.6 & 67.0 & 4.1 \\
\hline
3901254 & 106.9 & 8.7 & 51.7 & 11.3 & 70.1 & 7.7 \\
\hline
3901339 & 130.0 & 16.6 & 49.9 & 7.1 & 76.6 & 10.1\\
\hline
3903867 & 127.2 & 19.2 & 65.1 & 9.3 & 85.8 & 12.5\\
\hline
\bf Average & \bf 115.5 & \bf12.2 & \bf59.0 & \bf5.9 & \bf77.8 & \bf7.4 \\
\hline
\end{tabular}
\end{table}

\begin{table*}[!htbp]
\begin{threeparttable}  
{
\caption{Performance of our BPM solution for each subject used in this paper}
\centering
\begin{tabular}{|c||c|c|c||c|c|c||c|c|c|}

\hline
&  \multicolumn{3}{c||}{SBP } & \multicolumn{3}{c||}{DBP } & \multicolumn{3}{c|}{MAP} \\
\hline
Subject  & MAE(mmHg) & RMSD(mmHg) & r & MAE(mmHg) & RMSD(mmHg) & r & MAE(mmHg) & RMSD(mmHg) & r \\
\hline\hline
\cellcolor{red}3000397 & 3.1035 & 4.9188 & 0.868 & 1.5632 & 2.2002 & 0.731 & 1.9770 & 2.7603 & 0.846 \\
\hline
\cellcolor{orange(webcolor)}3001689 & 4.6059 & 7.0163 & 0.791 & 2.2535 & 3.6053 & 0.641 & 2.7295 & 4.0355 & 0.731\\
\hline
\cellcolor{yellow}3002090 & 3.3773 & 5.5152 & 0.517 & 1.9898 & 2.9113 & 0.647 & 2.1320 & 3.0707 & 0.639\\
\hline
\cellcolor{ao(english)}3006202 & 3.5003 & 5.9724 & 0.919 & 1.1111 & 1.9774 & 0.875 & 1.8128 & 3.1057 & 0.913\\
\hline
\cellcolor{ao}3011085 & 1.9821 & 3.5556 & 0.830 & 1.5181 & 2.5915 & 0.731 & 1.5641 & 2.6326 & 0.788\\
\hline
\cellcolor{aqua}3011536 & 1.9295 & 2.8839 & 0.947 & 1.3703 & 2.0418 & 0.883 & 1.5271 & 2.2347 & 0.923\\
\hline
\cellcolor{patriarch}3505162 & 4.7560 & 8.4902 & 0.948 & 2.3787 & 4.1041 & 0.896 & 3.0222 & 5.1515 & 0.936\\
\hline
\cellcolor{pink}3505469 & 2.4428 & 3.5453 & 0.764 & 2.1836 & 3.1442 & 0.732 & 1.7056 & 2.6751 & 0.767 \\
\hline
\cellcolor{olive}3901254 & 3.6524 & 5.4622 & 0.785 & 2.9212 & 7.5890 & 0.745 & 2.7983 & 5.2421 & 0.744 \\
\hline
\cellcolor{brown(web)}3901339 & 4.1607 & 6.0990 & 0.930 & 1.8092 & 3.0223 & 0.905 & 2.5261 & 3.7886 & 0.926\\
\hline
\cellcolor{gray}3903867 & 4.1938 & 6.1832 & 0.947 & 2.0237 & 3.0729 & 0.945 & 2.5851 & 3.8609 & 0.951\\
\hline
\bf Average & \bf 3.4277 & \bf 5.4220 & \bf 0.840 & \bf 1.9202 & \bf 3.2964 & \bf 0.793 & \bf 2.2163 & \bf 3.5052 & \bf 0.833 \\
\hline
\end{tabular}
\begin{tablenotes}
    \item  MAE: Mean Absolute Error; RMSD: Root Mean Squared Deviation; r:  Pearson correlation coefficient.
\end{tablenotes}
}
\end{threeparttable}  
\end{table*}

\subsection{Pre-processing}
Without extracting features from the PPG signal, our preprocessing applied on the real embedded device is relatively simple, including reducing the effects of noise and outliers, heartbeat segmentation, and converting PPG segments to 160-element vectors. 

\begin{itemize}

\item \textit{Bandpass Filtering and Normalization}:

The bandpass filter is applied to remove noise and drift (0.4Hz-8Hz), implemented by an order 4 Inverse Chebyshev bandpass filter, which is typically implemented in real-time application. To make the best practices for ANN training, we then apply normalization on entire PPG signal for each subject with the following expression:
$PPG_n = (PPG - min(PPG))/(max(PPG) - min(PPG))$, where normalized PPG signal as $PPG_n$.

\item \textit{Peak Detection, Segmentation and Outliers Cleaning}:

To robustly detect periodical peaks, we set the minimum horizontal distance in samples between neighboring peaks to be larger than 30 (0.24s) to detect peaks of the entire PPG signal, followed by a simply segmentation. Outliers cleaning is implemented by removing segments larger than 160 samples or smaller than 30 samples. At the same time, if the signal amplitude was outside ± 2 STDs from the mean, that heartbeat was removed from the dataset. Nevertheless, for all datasets employed no more than 7.2\% of the data were removed for any subject, with an average and STD of 4.3 ± 1.5\%.

\item \textit{Converting to Vectors}:

The last step before feeding input vectors into ANN is to convert PPG segments with different sizes to 160-element vectors that contained one heartbeat of sampled PPG values followed by zeros until the last element.
\end{itemize}
\begin{figure}[!htbp]
    \includegraphics[width=0.35\textwidth]{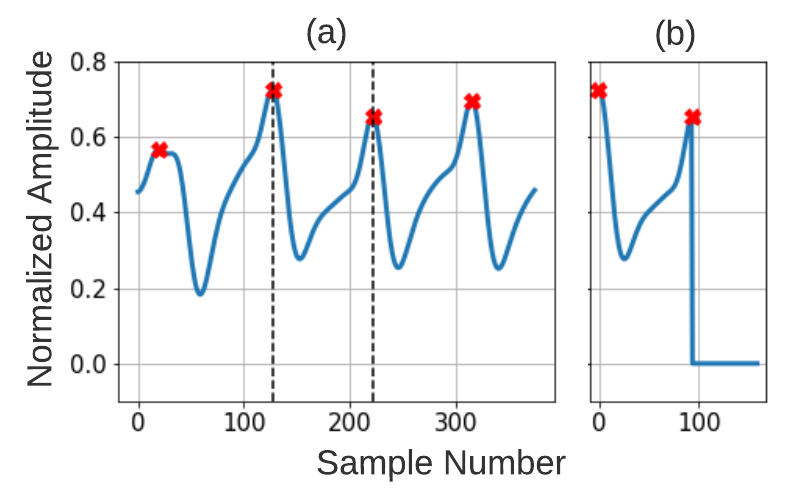}
    \centering
    \caption{PPG signal with preprocessing. a) Peak detection; b) Converting one heartbeat PPG segment (between dash lines) to a 160-element vector.}
\end{figure}

\subsection{ANN training}

The training is intra-patient based because pulse pattern differs from subject to subject, which can be calibrated by biometrics in future study. However, it is not in the scope of the paper. For each subject, we preprocess 10h signals into heartbeat samples labeled by BP values and randomly separate all data into two sets, 80\% for training and 20\% for testing. Compared to parameters extracted from the PPG waveform, our method feeds entire heartbeat PPG segments (extended to 160 samples) to the ANN as in Fig. 1. Three outputs, SBP, DBP, and MAP, are computed by using this feed-forward ANN with two hidden layers. The learning rate for training is tuned by callback functions. Training stop when the monitored loss metric has stopped improving within 20 epochs. The epoch number for different training processes is 114 ± 30 (mean ± STD) in our experiments, compared to the interval training epoch of ANN with PPG inputs 123 ± 100 in [3].

\section{Experimental Setup and Evaluation}

\subsection{Hardware Platform} 
Our solution is designed for low-power, low-memory wearable devices. Therefore, we choose a hardware platform equipped with an ultra-low-power EFM32 Leopard Gecko ARM Cortex-M3 32-bit microcontroller with a maximum clock rate of 48 MHz. It has 32 kB RAM, 256 kB Flash. Our solution applies to any wearable device having the above or similar specifications.

\subsection{Performance Evaluation}
We carry out intra-patient experiments to evaluate the performance of the proposed solution with segments of the same subject fed in the same training process. The experiment results of BPM performance are shown in Table II. Mean Absolute Error (MAE), Root Mean Squared Deviation (RMSD), and Pearson correlation coefficient (r) between the estimation values and ABP-extracted BP values (SBP, DBP, and MAP) are used as metrics to judge model performance. Average MAE and RMSD are 3.42 ± 5.42 mmHg for SBP, 1.92 ± 3.29 mmHg for DBP, and 2.21 ± 3.50 mmHg for MAP. Furthermore, the average mean error (ME) is -0.06 mmHg, -0.18 mmHg, -0.09 mmHg, and the average standard deviation of absolute error (STD) is 4.19 mmHg, 2.65 mmHg, 2.70 mmHg, respectively for SBP, DBP, and MAP. Based on the criteria for fulfilling the Association for the Advancement of Medical Instrumentation (AAMI) protocol [11], we can tell that the precision of our performance satisfies the AAMI criteria (RMSD $<$ 8 mmHg).

\begin{figure}[ht]
    \centering
    \includegraphics[width=0.38\textwidth]{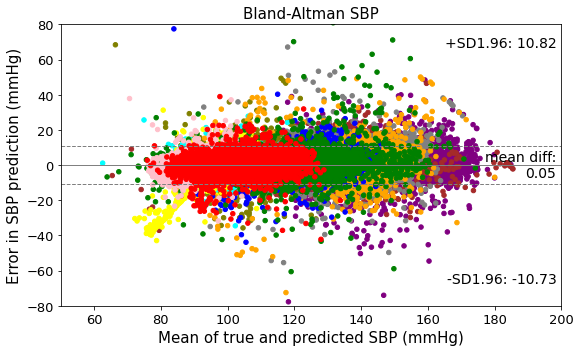}
    \includegraphics[width=0.38\textwidth]{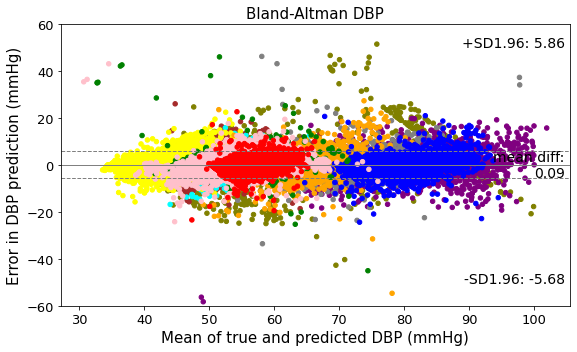}
    \caption{Bland-Altman plots of SBP and DBP. Subject color is defined in Table II.}
\end{figure}

Compared to the parameter-based method provide by [8] with 2-layer ANN (35 neurons on the first layer and 20 neurons on the second) which provide performance as 3.80 ± 3.46 mmHg (MAE ± STD) and 2.21 ± 2.09 mmHg for SBP and DBP respectively, our solution provides similar performance and has advantages of no feature extraction process and less ANN neurons.

Compared to the whole-based method provide by [6] which shows performance as 3.97 ± 8.9 (MAE ± RMSD) mmHg, 2.41 ± 4.18 mmHg, 2.61 ± 4.92 mmHg for SBP, DBP, and MAP respectively, our performance for SBP and DBP are better. However, they have 15s records extracted from hundreds of individuals, and our solution is most focused on 10h records from each subject. In our subject-based solution, personalization of models is important and substantially improves the results while deriving a good general predictive model is difficult.
As targeting low processing power wearable devices, the performance of continuous SBP estimation in [12] is -0.043 ± 6.79 mmHg (ME ± STD) with a 60Hz sampling rate. Compared to that, our solution has been evaluated on a platform with low power edge computing and has better performance. However, we evaluate the performance on a database with recordings at a 125Hz sampling rate.

\subsection{Energy consumption evaluation}
Energy profiling is done using the Simplicity studio software that comes with the EFM32 microcontrollers. As we have not seen any study provide the energy consumption of BPM solution, we provide execution time, average current, average power, and the energy calculated in Table III for future comparison. The C code is implemented for all the preprocessing steps and a pretrained ANN model is also implemented to do regression work of BP values. We have 3s of raw PPG feed into the Flash each time and evaluate the energy profiling based on the 375-element input. Our preprocessing step and ANN consume 1.14 mJ and 1.05 mJ respectively. Moreover, our preprocessing step executes with 18.1 KB of RAM and 27.0 KB of Flash, and the ANN step takes 0.1 KB of RAM and 6.1 KB of Flash. Thus, this solution would leave enough space for other applications to run parallel on wearable devices.

\begin{table}[!htbp]
\caption{energy consumption analysis on real hardware}
\centering
\begin{tabular}{>{\centering}b{0.11\textwidth}||>{\centering}b{0.06\textwidth}|>{\centering}b{0.06\textwidth}|>{\centering}b{0.07\textwidth}|>{\centering\arraybackslash}b{0.06\textwidth}}
\hline
 Our solution & Exe. Time(ms) &  Avg. Curr.(mA)  & Avg. Power(mW) & Energy (mJ)\\
\hline\hline
Preprocessing & 22.48 & 15.78 & 50.73 & 1.140\\
\hline
ANN  & 20.71 & 15.74 & 50.58 & 1.047 \\
\hline
\bf Entire solution  &\bf 42.25 &\bf 15.75 & \bf50.58 &\bf 2.137  \\
\hline
\end{tabular}
\end{table}

\section{DISCUSSION}

With our solution, waveform manipulation and BPM with low computational cost can be applied to wearable devices with precision that meets the AAMI requirement. We also proved our solution provides good performance and has the low-power and low memory property at the same time. We have made crucial parts of our research to provide a solid baseline for future study. However, there are several key problems to be solved in future work to make the solution be applied to more general scenarios.

First, the accuracy of the model deeply depends on the quality of the PPG pattern. Most wrist-worn wearables use green light to provide a good signal-noise ratio and robustness against motion artifacts [13]. At the same time, different strategies are applied to further suppress undesired noise like motion artifacts and to improve signal quality on wearables. Side-channel information for motion artifact cancellation (MAC) along with single-site PPG would be necessary to do BPM in exercising scenario, like an accelerometer [14] or multi-channel PPG with common-mode rejection [15]. To implement MAC efficiently with low computational cost would be one of the future focuses of our work.

Secondly, the database we used is with traditional uniform sampling at a fixed-sampling frequency. Different strategies are used on commercial wearables to achieve a good signal while using a minimum of energy [13]. Based on the need for different qualities of readings, sampling strategies could be applied to help reduce monitoring energy while having acceptable BPM accuracy. 

Finally, it is well-known that calibration is also one of the main challenges in cuffless BPM. The initial calibration against a cuff-based device can hardly be avoided. The calibration procedure can extremely increase the estimation accuracy of BP because calibration provides a reference BP level for patients [9]. To overcome the calibration limitation, biometrics like age, height, weight, and BP levels shows great potential to reduce fitting errors and calculate BP in certain populations [5]. Therefore, based on the profiles that come with wearable devices, BPM based on PPG would benefit to become calibration-free in certain populations. 
   
\section{CONCLUSIONS}
This paper provides an energy-efficient solution for continuous and non-invasive BPM based on single-site PPG signal and ANN with edge computing. The performance of this solution shows good accuracy compared to other studies. The main contributions of this paper are as follows:

\begin{itemize}
\item A real-time BPM solution with edge computing, which is fast, energy-efficient, low-memory consumed, and provides accurate BPM on wearable devices.
\item Performance validation of our solution on the MIMIC-III database.
\item Energy efficiency validation of our solution on real hardware platform.

\end{itemize}

\addtolength{\textheight}{-12cm}   


\end{document}